\documentclass[aps,pre,preprintnumbers,floatfix,nofootinbib,11pt]{revtex4-1}
\usepackage{amsmath}
\usepackage{amssymb}
\usepackage{graphicx,color}
\usepackage{multirow}
\usepackage[switch, modulo]{lineno}

\pdfoutput=1

\begin{document}

\title{A connection between the Ice-type model of Linus Pauling and the
three-color problem}
\author{Roberto da Silva$^{1}$, Osvaldo S. Nakao$^{2}$, J. R. Drugowich de
Fel\'{\i}cio$^{3}$}

\address{1 - Instituto de F\'{i}sica, Universidade Federal do Rio Grande do Sul, Porto Alegre, Rio Grande do Sul, Brazil\\
	2 -  Departamento de Engenharia de Estruturas e Geot\'{e}cnica, Escola Polit\'{e}cnica da Universidade de S\~{a}o Paulo, S\~{a}o Paulo, S\~{a}o Paulo, Brazil\\
	3 - Departamento de F\'{i}sica, Faculdade de Filosofia, Ci\^{e}ncias e Letras de Riber\~{a}o Preto, Universidade de S\~{a}o Paulo, Ribeir\~{a}o Preto, S\~{a}o Paulo, Brazil}

\begin{abstract}
The ice-type model proposed by Linus Pauling to explain its entropy at low
temperatures is here approached in a didactic way. We first present a
theoretically estimated low-temperature entropy and compare it with
numerical results. Then, we consider the mapping between this model and the
three-colour problem, i.e.,colouring a regular graph with coordination equal
to 4 (a two-dimensional lattice) with three colours, for which we apply the
transfer-matrix method to calculate all allowed configurations for
two-dimensional square lattices of $N$ oxygen atoms ranging from 4 to 225.
Finally, from a linear regression of the transfer matrix results, we obtain
an estimate for the case $N\rightarrow \infty $ which is compared with the
exact solution by Lieb.
\end{abstract}

\maketitle

\section{Introduction}

\label{sec:introduction}

Counting problems are part of our daily lives, normally disguised in the
format of probability calculation. There is no way to estimate probabilities
(at least within the Laplace definition) without enumerating all
possibilities (the sample space), which is why lottery apportionment is
frequently postponed to later weeks. In addition to challenging gamblers,
this type of games often resists the cleverness and intelligence of
mathematicians, statisticians, and physicists. This is what happened with
the so-called ice-type model, the subject of this article. It can be
summarized with a simple question: How many different ways can hydrogen
bonds be arranged\ if water is frozen to $T=0$\ K?\ In the nomenclature of
thermodynamics, what is the residual entropy of the ice, i.e. $S=k_{B}\ln
\Omega $? Here, $k_{B}$\ is the Boltzmann constant, $\Omega $\ is the number
of accessible configurations to the system with $N$\ oxygen atoms. This is
not only a curiosity. Such problem appeared in 1933 when Giauque \& Ashley 
\cite{Giauque} measured the entropy of ice at low temperatures and found the
molar entropy to be $s=0.82\pm 0.05\ $cal/(mol\ K), remembering that $1$cal$%
\ \approx 4.18$ J. The molar entropy corresponds to the product of the
Avogrado number ($N_{0}$) by the entropy per site, which is $S/N=k_{B}\frac{1%
}{N}\ln \Omega =k_{B}\ln \Omega ^{1/N}$ or $k_{B}\ln W$, where $W=\Omega
^{1/N}$\textbf{.}

It is important to note that the answer depends on the spatial dimension in
which the H$_{2}$O molecules are inserted.

The first theoretical estimate for the residual entropy of ice was published
in 1935 by Linus Pauling who obtained $s=0.805$\ cal/(mol K) \cite{Pauling}.
This value is in good agreement with the experimental value, despite the
several approximations used by Pauling. In Lieb's own words \cite{Lieb},
"this calculation must be considered as one of the more fortunate
applications of statistical mechanics to real substances". Only in the
1960's Nagle \cite{Nagle} performed more precise numerical estimates for the
entropy for a three-dimensional lattice and taking into account interaction
among the vertices in a structure that simulates the ice and obtained $%
s=0.8145\pm 0.0002$\ cal/(mol\ K). Both results are in complete agreement
with the experimental estimate obtained by Giauque and Ashley \cite{Giauque}.

Considering a two-dimensional version for the ice, Nagle obtained $s=$\ $%
0.8580\pm 0.0013\ $cal/(mol\ K) \cite{Nagle,Nagle2}. Lieb \cite{Lieb}
obtained an exact solution for the problem in two dimensions ($s=0.856$\
cal/(mol K)).

It is important to mention that the ice has a tetrahedral three-dimensional
structure with oxygen atoms occupying the vertices. Thus, each oxygen atom
is linked by hydrogen bonds to four other oxygen atoms. This means that the
two-dimensional version of the model preserves a fundamental characteristic
of the real structure which is the fact of each oxygen atom has four
neighbours, with hydrogen atoms between them. Since each water molecule has
only two hydrogen atoms, in the real structure, two of these hydrogen atoms
must be in the nearest equilibrium position ($d=0.95$ \AA ) and the two
other at the larger distance ($d=1.81$ \AA ), which belong to the
neighbouring oxygen atoms. We shall come back to this subject and show that
it corresponds to consider neutral molecules, giving rise to the well-known
six-vertex model in Statistical Mechanics.

As the note added in proof of the paper by Lieb due to an observation of A.
Lennard \cite{Lieb}, this problem is equivalent, except by a factor 3, to
discover how many ways exist to paint a square map using only 3 colours. A
proper colouring of a map means that two neighbour countries cannot have the
same colour, for if two countries have the same colour the border between
them would disappear, i.e., the two countries would be merged into one.

In the next section we introduce the ice-type model and reproduce the
calculation presented by Linus Pauling. In section \ref{Section: colors}, we
show the equivalence between this problem and the three-colouring map. In
the same section, we also enumerate the acceptable configurations for small
systems and present the numerical solution of the problem using a
\textquotedblleft brute force\textquotedblright\ method motivating our next
section.

In section \ref{Section: Numerical_and_transfer} we present an original
numerical calculation developed for the colouring version of the problem
which allows to proceed to the systems with $N=225$ atoms. In addition,
after an extrapolation to the thermodynamic limit ($N\rightarrow \infty $),
our numerical estimate is compared with the exact result by Lieb \cite{Lieb}%
. Finally, we present some conclusions in section \ref{Section: conclusions}.

\section{The estimate of Linus Pauling}

\label{Section: Estimate_linus_pauling}

\begin{figure}[h]
\begin{center}
\includegraphics[width=1.0\columnwidth]{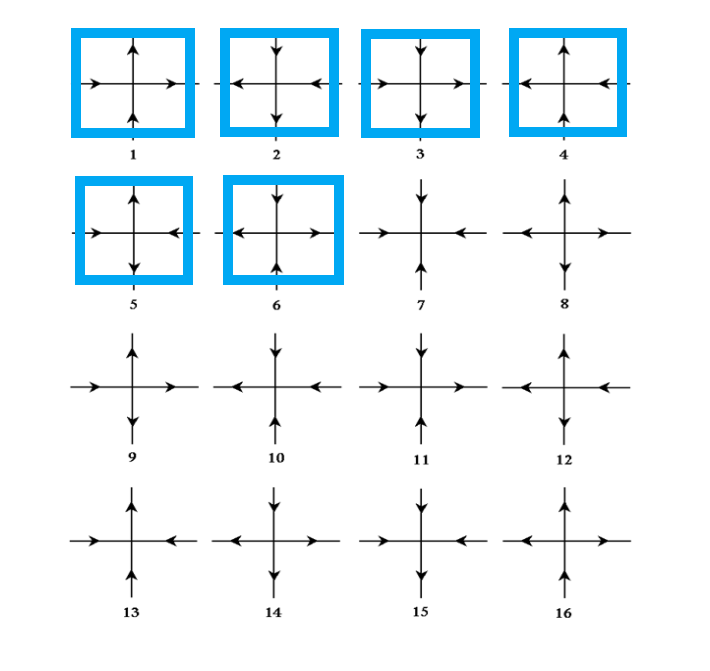}
\end{center}
\caption{The sixteen possibilities of the vertex-types with only the six
first vertices satisfying the ice rules, i.e., satisfying the conservation
law where two and only two arrows are coming in (which means two and only
two arrows coming out) (the allowed vertices outlined in blue).}
\label{Fig:ice_type_arrows}
\end{figure}

Since each hydrogen atom can be in two distinct positions, Pauling used an
arrow to indicate if it is near (arrow comes in) or far (arrow exits of)
each oxygen atom. The percentages of ions $H_{3}O^{+}$ and $OH^{-}$ are
taken as zero which means that each oxygen atom (site in the lattice) must
necessarily have 2 and only 2 hydrogen atoms next to it (two arrows arriving
and two arrows departing from a site). The consequence is that from 16
possible types of vertices in Fig. \ref{Fig:ice_type_arrows}, only $6$ (the
first six vertices in that same plot, outlined in blue) satisfy the
so-called \textquotedblleft ice rules\textquotedblright\ introduced by
Bernal and Fowler \cite{Bernal}\ in 1933 and improved by Pauling.

This problem gave rise to the so called six-vertex model which was solved
exactly in the 1960s. Returning to Pauling's calculation, there is a
fraction of $6/16$ allowed vertices in each site and the total number of
configurations of the lattice (when one randomly chooses the direction of
the $2N$ arrows starting from the $N$ oxygen atoms) is $2^{2N}$. Thus,
assuming statistically independent sites (which is not true) and that in all 
$N$ vertices the ice rules are satisfied, we have 
\begin{equation}
\begin{array}{lll}
\Omega & \cong & 2^{2N}(\frac{6}{16})^{N}%
\end{array}%
\end{equation}%
or 
\begin{equation*}
W\cong \dfrac{3}{2},
\end{equation*}%
which leads to an entropy per mol equal to

\begin{equation*}
s=N_{0}k_{B}\ln W\cong R\ln (1.5)=0.805\ \text{cal}/(\text{mol\ }K),
\end{equation*}%
where $N_{0}$ is the Avogadro number, $k_{B}$ is the Boltzmann constant and $%
\ R=N_{0}k_{B}=1.985$ cal$/($mol\ $K)$ \ is the universal gas constant.

\section{The language of colours}

\label{Section: colors}

In this section we will show how the six-vertex model can be mapped onto the
three colour problem.

\begin{figure}[h]
\begin{center}
\includegraphics[width=1.0\columnwidth]{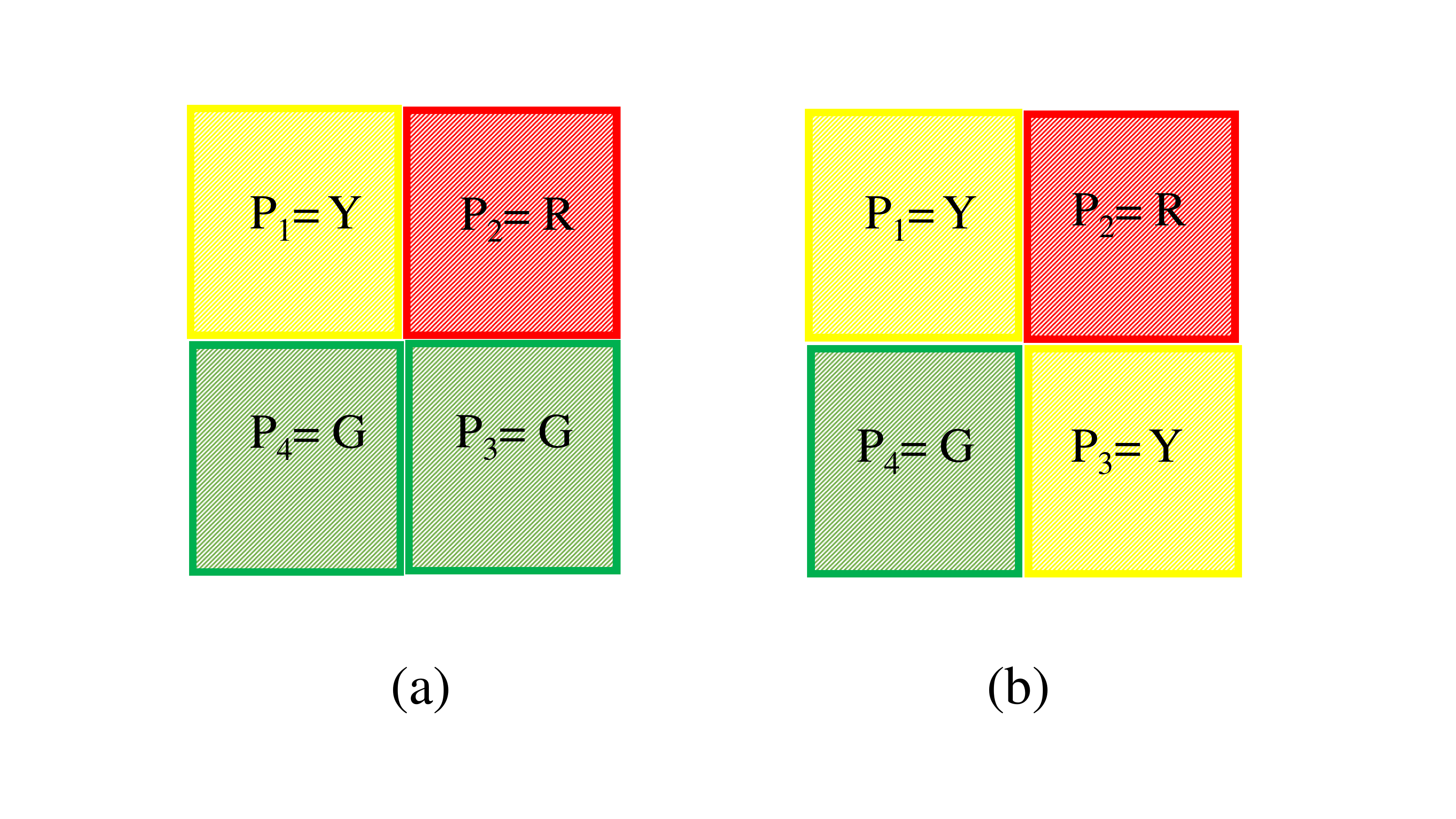} %
\includegraphics[width=1.0\columnwidth]{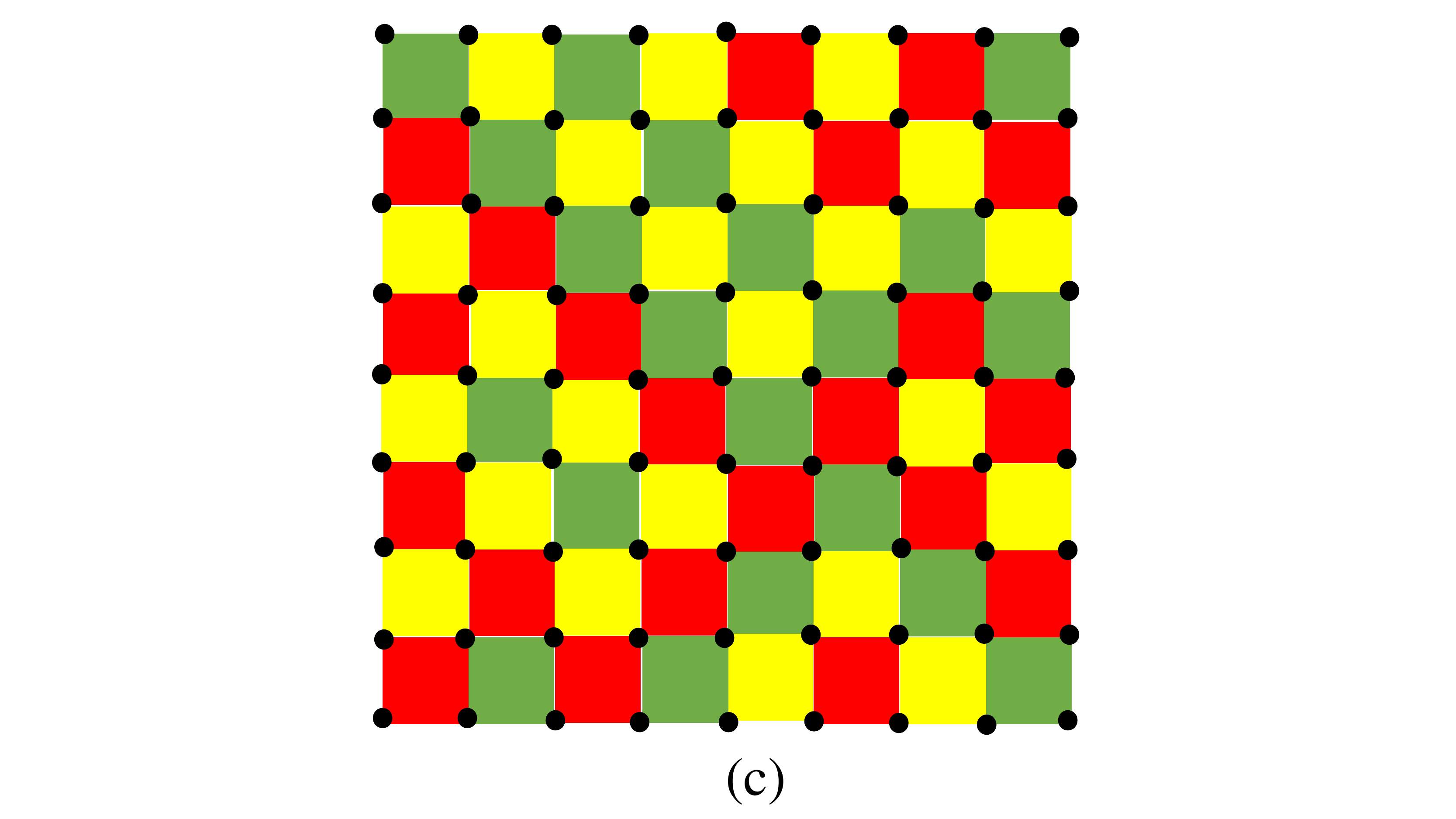}
\end{center}
\caption{(a) Improper colouring (b) Proper Colouring. (c) Example of a
proper colouring of a map with $8\times 8$ countries.}
\label{Fig:coloring}
\end{figure}

Let us familiarize with the problem of colouring maps by analysing a small
\textquotedblleft world\textquotedblright with only 4 countries. There are
81 ways of painting the map with three colours\footnote{%
One may think that the correct value would be $4^{3}\ $but this is not true.
Using the analogy of rolling two dices, the total number of different
outcomes is $6^{2}=36$ and not $2^{6}=64$. Each of the six faces of the
first dice can appear together with any of the six faces of the second dice.
Therefore, one has $6\times 6$ $=6^{2}$ possibilities. With three dices, one
has $6\times 6\times 6$ $=6^{3}$ possibilities. Thus, the correct is to take
as basis the number of possible states for each entity: (dice face, country
colour, etc.) and as exponent the number of entities (number of dices,
number of countries, etc.)}, resulting from the operation with the product $%
(3\times 3\times 3\times 3)$ = $3^{4}$ since each country can be painted (in
principle) with any of the three colours. However, many of these 81 ways do
not satisfy the condition according to which two adjacent countries cannot
be painted with the same colour. This condition destroys the independence
between events, thus prohibiting the multiplication $3\times 3\times 3\times
3=3^{4}$ and drastically reducing the number of acceptable paintings. There
is no simple reasoning to calculate this number because the events are not
independent, but we may nevertheless solve the problem for a small number of
countries. Let us consider that the countries are coloured with yellow (Y),
green (G), and red (R). Fig. \ref{Fig:coloring} shows the countries labelled
as $P_{1},P_{2},P_{3}$ e $P_{4}$, starting from the upper left corner and
rotating clockwise. The arrangement in Fig. \ref{Fig:coloring} (a) does not
satisfy the constraint as two neighbour countries are painted with the same
colour ($P_{3}=P_{4}$). In contrast, the constraint is fulfilled in Fig \ref%
{Fig:coloring} (b) where all neighbours are painted with distinct colours.

We can now enumerate the number of distinct paintings keeping in mind that
after painting one of the countries, its neighbours can be coloured with one
of the two remaining colours. For instance, if yellow is chosen to paint
country $P_{1}$, country $P_{2}$ can only be painted with colours red or
green. In addition, country $P_{4}$ has to satisfy the same constraint which
means that choosing the colour of country $P_{1}$ one has $2+2=4$ ways to
paint countries $P_{1}$, $P_{2}$ and $P_{4}$. For the remaining country $%
P_{3}$, there are the following alternatives: Either its two neighbours ($%
P_{2}$ and $P_{4}$) are painted with the same colour (both with the colour
red or both with the colour green) or they are painted with different
colours. In the first case country $P_{3}$ can be painted with colour
yellow, the same colour of $P_{1}$, or with a remaining colour (G or R).
Therefore, the two first possibilities are multiplied by 2, leading to 4
possible configurations. In the second case ($P_{2}$ and $P_{4}$ with
different colours) there is no other way to paint $P_{3}$: it must be
painted with the colour yellow (the same colour of $P_{1}$). The total
number is then 6, the same found by Pauling for the number of vertices that
satisfy the iced-type rules. However this is not yet the final result for
the case of colours. We obtained 6 paintings by choosing yellow to paint $%
P_{1}$. There are 6 more possible colourings starting with Red for $P_{1}$
and another 6 by painting the first country with green. Therefore, the total
number is $3\times 6=18$ different ways to colour a 4-country world with 3
colours.

But, why is this problem interesting to us? How does it work for larger
worlds? In the next subsection we will show that the 3-colouring problem is
equivalent to the six-vertex problem except by a multiplicative factor. It
is important to notice that the colouring can contemplate more complex
structures than simply maps (regular graphs) and using a different number of
colours which is known graph-colouring. For the interested readers we
strongly suggest read the Appendix I. In this appendix we present the
problem in the light of the graph theory by showing extensions. In addition
a report about analytical results related to the 3-colouring problem.

\subsection{Mapping the six-vertex problem onto the three-colouring problem}

\label{Subsection:Mapping}

Let us attribute the same three colours (yellow, green, and red) to
countries in larger maps with the restriction that neighbouring countries
cannot have the same colour. Fig \ref{Fig:coloring} (c) shows a example of a
proper colouring of a map with $8\times 8$ countries. .

We shall use a cyclical convention for the colours: Y follows G, G follows
R, and R follows Y (YGRYGRYGR...). Every time that, rotating clockwise in
relation to a perpendicular axis to the lattice plane, passing by the common
point to the neighbouring 4 countries (the black points in Fig \ref%
{Fig:coloring} (c)), and let us considering all worlds of the 4 countries in
the $L\times L$ worlds.

Starting for example by convention from $P_{1}$ in each of these small
worlds of the four countries, if we change from yellow to green (or from
green to red or from red to yellow) the arrow in the boundary will be
directed to the common point (blue arrow), while the changing from red to
green (or from green to yellow or from yellow to red) it will be exiting
from the common point (orange arrow), and this for both situations:
horizontal and vertical boundaries.

\begin{figure}[h]
\begin{center}
\includegraphics[width=1.0\columnwidth]{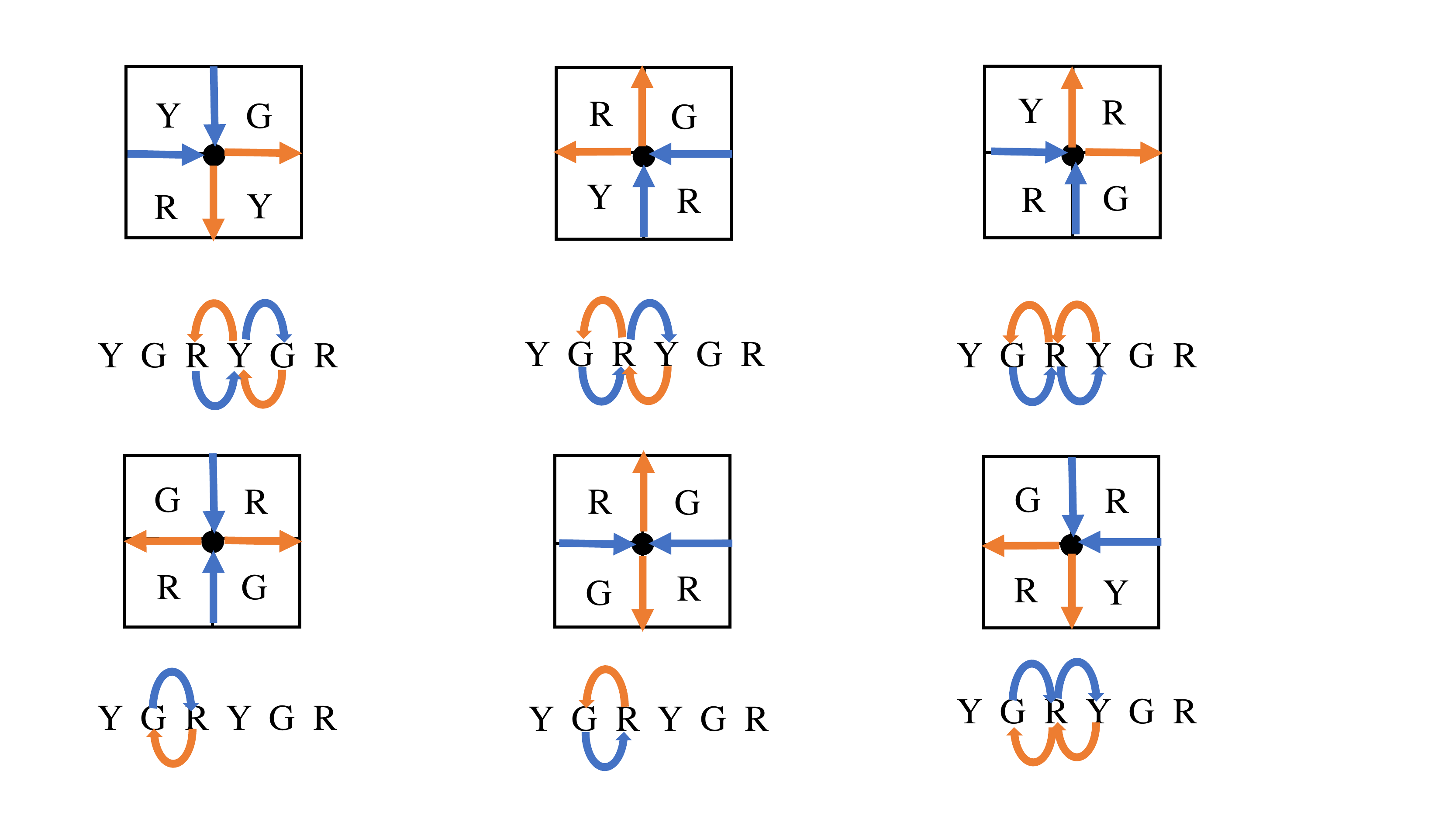}
\end{center}
\caption{Examples of colourings of worlds with four countries translated to
the possible 6 configurations of arrows. We start from P1 and rotate
clockwise until P1 again following the cyclical convention
established,considering blue (arrow in) and orange (arrow out). Any other
configuration of colours is translated into one of these possible 6
configurations of arrows.}
\label{Fig:countries_and_graphs_3}
\end{figure}

This common point that resembles the position of oxygen atoms will always
have two arrows in and two arrows out. Fig. \ref{Fig:countries_and_graphs_3}
shows the six possible configurations of the arrows.

Let us consider the first colour configuration in that same figure \ref%
{Fig:countries_and_graphs_3}. We start in $P_{1}$ with Y and $P_{2}$ with G,
and following the clockwise orientation there is a blue vertical arrow
pointing to the black point. From $P_{2}$ to $P_{3}$, one has G to Y,
following the counter-clockwise direction, and thus there is an orange
horizontal arrow out from the black point. From $P_{3}$ to $P_{4}\ $ there
is an orange vertical arrow out of the black point, and from $P_{4}$ to $%
P_{1}$ there is a blue horizontal arrow pointing to the black point. The
other colourings in this translation only can lead to the one of possible
arrow configurations represented in Fig. \ref{Fig:countries_and_graphs_3}

It is reminded that each configuration of arrows corresponds to three
possible colourings since there are three possible colours to start in $%
P_{1} $.

Thus we can write 
\begin{equation}
\Omega _{\text{colours}}(N)=3W^{N}  \label{Eq:relation_omega_e_W}
\end{equation}%
and the factor 3 links the number of configurations of six-vertex problem to
the three-colour problem.

However to solve this problem using a computer should be interesting to know
if we can perform this calculation for larger values of $N$. In the next
section we start this task first using an algorithm which calculates $\Omega
_{\text{colours}}(N)$ using brute force (BF).

\section{Computers and Brute Force: Preliminary Numerical Results}

A table can be generated by showing the increasing of the number of
possibilities $\Omega _{\text{colours}}(N)$ when the number of countries
increases, which can be done by enumerating all possibilities of painting
discarding the ones that do not satisfy the condition of the problem. A
Fortran program (with Fortran 77), which can compiled in any Fortran
compiler, with a few lines is sufficient to calculate $\Omega _{\text{colours%
}}(N)$ for a world with four countries (see Table: \ref{Table:Brutal_force}%
). This is carried out using \textquotedblleft brute force\textquotedblright
, i.e., checking all possible colourings in the map. In these programs we
consider that G becomes 0, R becomes 1, and Y becomes 2.

%TCIMACRO{\TeXButton{B}{\begin{table}[tbp] \centering}}%
%BeginExpansion
\begin{table}[tbp] \centering%
%EndExpansion
\begin{tabular}{ll}
\hline\hline
& \textbf{Program}: Colouring a small world with 4 countries using exaustive
enumeration (\textquotedblleft Brute Force \textquotedblright) \\ 
0 & Integer $P_{1},P_{2},P_{3},P_{4},Icount$ \\ 
1 & $Icount=0$ \\ 
2 & \ \ \ \ \ \ Do $P_{1}=0,2$ \\ 
3 & \ \ \ \ \ \ \ \ \ \ \ Do $P_{2}=0,2$ \\ 
4 & \ \ \ \ \ \ \ \ \ \ \ \ \ \ Do $P_{3}=0,2$ \\ 
5 & \ \ \ \ \ \ \ \ \ \ \ \ \ \ \ \ \ Do $P_{4}=0,2$ \\ 
6 & \ \ \ \ \ \ \ \ \ \ \ \ \ \ \ \ \ \ \ \ \ If ($P_{1}$.ne.$P_{2}$.and.$%
P_{1}$.ne.$P_{4}$) then \\ 
7 & \ \ \ \ \ \ \ \ \ \ \ \ \ \ \ \ \ \ \ \ \ \ \ \ \ If($P_{2}$.ne.$P_{3}$%
.and.$P_{4}$.ne.$P_{3}$) then \\ 
8 & $\ \ \ \ \ \ \ \ \ \ \ \ \ \ \ \ \ \ \ \ \ \ \ \ \ \ \ \ \
Icount=Icount+1$ \\ 
9 & \ \ \ \ \ \ \ \ \ \ \ \ \ \ \ \ \ \ \ \ \ \ \ \ \ Endif \\ 
10 & \ \ \ \ \ \ \ \ \ \ \ \ \ \ \ \ \ \ \ \ \ Endif \\ 
11 & \ \ \ \ \ \ \ \ \ \ \ \ \ \ \ \ Enddo \\ 
12 & \ \ \ \ \ \ \ \ \ \ \ \ Enddo \ \ \ \ \ \ \  \\ 
13 & \ \ \ \ \ \ \ \ \ Enddo \\ 
14 & \ \ \ \ \ \ Enddo \\ 
15 & Write(*,*)$%
%TCIMACRO{\U{b4}}%
%BeginExpansion
{\acute{}}%
%EndExpansion
Number\ of\ configurations$ =$%
%TCIMACRO{\U{b4}}%
%BeginExpansion
{\acute{}}%
%EndExpansion
$,$\ Icount$ \\ 
16 & Stop \\ 
17 & End \\ \hline\hline
\end{tabular}%
\caption{A Brute Force algorithm}\label{Table:Brutal_force}%
%TCIMACRO{\TeXButton{E}{\end{table}}}%
%BeginExpansion
\end{table}%
%EndExpansion

\begin{figure}[h]
\begin{center}
\includegraphics[width=1.0\columnwidth]{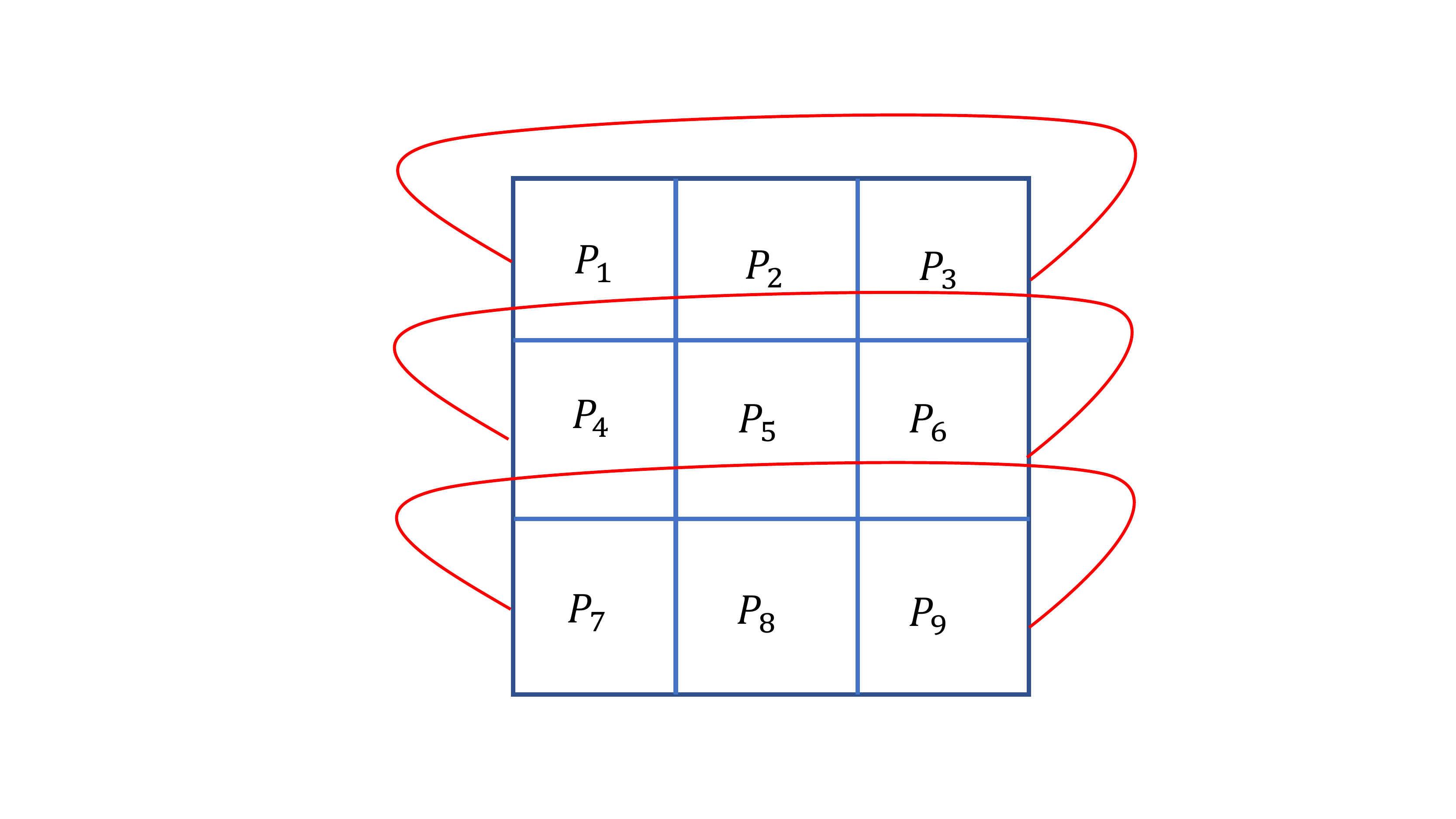}
\end{center}
\caption{Configuration with $N=9$ countries. The red lines correspond to
neighbourhood relations created for periodic boundary conditions. Such
connections must be removed for free boundary conditions.}
\label{Fig:PBC_and_FBC}
\end{figure}

There is no increase in difficulty to replace a world of 4 countries to
another with 9. The program will enumerate $3^{9}=\allowbreak 1.\,9683\times
10^{4}$ configurations for $9$ countries, using one of two alternatives:
Free boundary conditions (FBC) or Periodic boundary conditions (PBC) in one
of the directions to perform the counting. This was not important for 4
countries because in that case there was no difference between FBC and PBC.
However, for $N=9$ countries this is not the case. For example, $P_{1}$ is
neighbour to $P_{3}$, or $P_{7}$ is neighbour to $P_{9}$ with PBC but such
neighbourhood relations are not considered for FBC, i.e., the red links in
Fig. \ref{Fig:PBC_and_FBC} are removed. Note, for instance, how the brute
force algorithm for $N=9$ with PBC (Table \ref{Table:Brutal_force_Neq9})
demands a lot more of "brute force" compared to the case of $N=4$.

%TCIMACRO{\TeXButton{B}{\begin{table}[tbp] \centering}}%
%BeginExpansion
\begin{table}[tbp] \centering%
%EndExpansion
\begin{tabular}{ll}
\hline\hline
& \textbf{Program}: Colouring \textquotedblleft Brute Force $N=9$%
\textquotedblright \\ 
0 & Integer $%
P_{1},P_{2},P_{2},P_{3},P_{4},P_{5},P_{6},P_{7},P_{8},P_{9},Icount$ \\ 
1 & $Icount=0$ \\ 
2 & \ \ \ \ \ \ Do $P_{1}=0,2$ \\ 
3 & \ \ \ \ \ \ \ \ \ \ \ Do $P_{2}=0,2$ \\ 
4 & \ \ \ \ \ \ \ \ \ \ \ \ \ \ Do $P_{3}=0,2$ \\ 
5 & \ \ \ \ \ \ \ \ \ \ \ \ \ \ \ \ \ Do $P_{4}=0,2$ \\ 
6 & \ \ \ \ \ \ \ \ \ \ \ \ \ \ \ \ \ \ \ \ \ Do $P_{5}=0,2$ \\ 
7 & \ \ \ \ \ \ \ \ \ \ \ \ \ \ \ \ \ \ \ \ \ \ \ \ \ Do $P_{6}=0,2$ \\ 
8 & \ \ \ \ \ \ \ \ \ \ \ \ \ \ \ \ \ \ \ \ \ \ \ \ \ \ \ \ \ \ \ Do $%
P_{7}=0,2$ \\ 
9 & \ \ \ \ \ \ \ \ \ \ \ \ \ \ \ \ \ \ \ \ \ \ \ \ \ \ \ \ \ \ \ \ \ \ \ \
\ Do $P_{8}=0,2$ \\ 
10 & \ \ \ \ \ \ \ \ \ \ \ \ \ \ \ \ \ \ \ \ \ \ \ \ \ \ \ \ \ \ \ \ \ \ \ \
\ \ \ \ \ \ Do $P_{9}=0,2$ \\ 
11 & \ \ \ \ \ \ \ \ \ \ \ \ \ \ \ \ \ \ \ \ \ \ \ \ \ \ \ \ \ \ \ \ \ \ \ \
\ \ \ \ \ \ \ \ \ \ \ \ \ \ \ If ($P_{1}$.ne.$P_{2}$.and.$P_{1}$.ne.$P_{4}$)
then \\ 
12 & \ \ \ \ \ \ \ \ \ \ \ \ \ \ \ \ \ \ \ \ \ \ \ \ \ \ \ \ \ \ \ \ \ \ \ \
\ \ \ \ \ \ \ \ \ \ \ \ \ \ \ \ \ \ \ \ If ($P_{2}$.ne.$P_{3}$.and.$P_{2}$%
.ne.$P_{5}$) then \\ 
13 & \ \ \ \ \ \ \ \ \ \ \ \ \ \ \ \ \ \ \ \ \ \ \ \ \ \ \ \ \ \ \ \ \ \ \ \
\ \ \ \ \ \ \ \ \ \ \ \ \ \ \ \ \ \ \ \ \ \ \ \ \ If ($P_{4}$.ne.$P_{5}$.and.%
$P_{4}$.ne.$P_{7}$) then \\ 
14 & \ \ \ \ \ \ \ \ \ \ \ \ \ \ \ \ \ \ \ \ \ \ \ \ \ \ \ \ \ \ \ \ \ \ \ \
\ \ \ \ \ \ \ \ \ \ \ \ \ \ \ \ \ \ \ \ \ \ \ \ \ \ \ \ If($P_{5}$.ne.$P_{6}$%
.and.$P_{5}$.ne.$P_{8}$) then \\ 
15 & \ \ \ \ \ \ \ \ \ \ \ \ \ \ \ \ \ \ \ \ \ \ \ \ \ \ \ \ \ \ \ \ \ \ \ \
\ \ \ \ \ \ \ \ \ \ \ \ \ \ \ \ \ \ \ \ \ \ \ \ \ \ \ \ \ \ \ \ If($P_{3}$%
.ne.$P_{6}$.and.$P_{6}$.ne.$P_{9}$.and.$P_{9}$.ne.$P_{8}$.and.$P_{8}.$ne.$%
P_{7}$) then \\ 
16 & ** Inclusion of the conditions for the PBC in one direction: \\ 
17 & \ \ \ \ \ \ \ \ \ \ \ \ \ \ \ \ \ \ \ \ \ \ \ \ \ \ \ \ \ \ \ \ \ \ \ \
\ \ \ \ \ \ \ \ \ \ \ \ \ \ \ \ \ \ \ \ \ \ \ \ \ \ \ \ \ \ \ \ \ \ \ If($%
P_{1}$.ne.$P_{3}$.and.$P_{4}$.ne.$P_{6}$.and.$P_{7}$.ne.$P_{9}$) then \\ 
18 & $\ \ \ \ \ \ \ \ \ \ \ \ \ \ \ \ \ \ \ \ \ \ \ \ \ \ \ \ \ \ \ \ \ \ \
\ \ \ \ \ \ \ \ \ \ \ \ \ \ \ \ \ \ \ \ \ \ \ \ \ \ \ \ \ \ \ \ \ \ \ \ \ \
\ \ \ \ \ \ \ \ \ Icount=Icount+1$ \\ 
19 & \ \ \ \ \ \ \ \ \ \ \ \ \ \ \ \ \ \ \ \ \ \ \ \ \ \ \ \ \ \ \ \ \ \ \ \
\ \ \ \ \ \ \ \ \ \ \ \ \ \ \ \ \ \ \ \ \ \ \ \ \ \ \ \ \ \ \ \ \ \ \ Endif
\\ 
20 & \ \ \ \ \ \ \ \ \ \ \ \ \ \ \ \ \ \ \ \ \ \ \ \ \ \ \ \ \ \ \ \ \ \ \ \
\ \ \ \ \ \ \ \ \ \ \ \ \ \ \ \ \ \ \ \ \ \ \ \ \ \ \ \ \ \ \ \ Endif \\ 
21 & \ \ \ \ \ \ \ \ \ \ \ \ \ \ \ \ \ \ \ \ \ \ \ \ \ \ \ \ \ \ \ \ \ \ \ \
\ \ \ \ \ \ \ \ \ \ \ \ \ \ \ \ \ \ \ \ \ \ \ \ \ \ \ \ Endif \\ 
22 & \ \ \ \ \ \ \ \ \ \ \ \ \ \ \ \ \ \ \ \ \ \ \ \ \ \ \ \ \ \ \ \ \ \ \ \
\ \ \ \ \ \ \ \ \ \ \ \ \ \ \ \ \ \ \ \ \ \ \ \ Endif \\ 
23 & \ \ \ \ \ \ \ \ \ \ \ \ \ \ \ \ \ \ \ \ \ \ \ \ \ \ \ \ \ \ \ \ \ \ \ \
\ \ \ \ \ \ \ \ \ \ \ \ \ \ \ \ \ \ \ \ Endif \\ 
24 & \ \ \ \ \ \ \ \ \ \ \ \ \ \ \ \ \ \ \ \ \ \ \ \ \ \ \ \ \ \ \ \ \ \ \ \
\ \ \ \ \ \ \ \ \ \ \ \ \ \ \ Endif \\ 
25 & \ \ \ \ \ \ \ \ \ \ \ \ \ \ \ \ \ \ \ \ \ \ \ \ \ \ \ \ \ \ \ \ \ \ \ \
\ \ Enddo \\ 
26 & \ \ \ \ \ \ \ \ \ \ \ \ \ \ \ \ \ \ \ \ \ \ \ \ \ \ \ \ \ \ \ Enddo \ \
\ \ \ \ \  \\ 
27 & \ \ \ \ \ \ \ \ \ \ \ \ \ \ \ \ \ \ \ \ \ \ \ \ \ \ Enddo \\ 
28 & \ \ \ \ \ \ \ \ \ \ \ \ \ \ \ \ \ \ \ \ \ \ Enddo \\ 
29 & \ \ \ \ \ \ \ \ \ \ \ \ \ \ \ \ \ Enddo \\ 
30 & \ \ \ \ \ \ \ \ \ \ \ \ \ Enddo \\ 
31 & \ \ \ \ \ \ \ \ \ Enddo \\ 
32 & \ \ \ \ \ \ Enddo \\ 
33 & \ \ \ Enddo \\ 
34 & Write(*,*)$%
%TCIMACRO{\U{b4}}%
%BeginExpansion
{\acute{}}%
%EndExpansion
Number\ of\ configurations$ =$%
%TCIMACRO{\U{b4}}%
%BeginExpansion
{\acute{}}%
%EndExpansion
$,$\ Icount$ \\ 
35 & Stop \\ 
36 & End \\ \hline\hline
\end{tabular}%
\caption{"Brute Force" algorithm for N=9}\label{Table:Brutal_force_Neq9}%
%TCIMACRO{\TeXButton{E}{\end{table}}}%
%BeginExpansion
\end{table}%
%EndExpansion

For $N=16$ there are $3^{16}=\allowbreak 4.\,30467\,21\times 10^{7}$ colour
configurations to select among them the acceptable colourings, which
increase to $3^{25}$ or $8.\,47288\,60944\,3\times 10^{11}$ for $N=25$.

And how long would it take a personal computer to generate all those
configurations? Well, it depends on the time it takes to generate each
configuration. Actually, nowadays we have very fast personal computers and
they can execute small operations in thousandths of billionths of a second.
Technically, the computational performance is measured in FLOPS
(floating-point operations per second) and so actually the computers operate
in the scale of GIGAFLOPS and beyond.

For $N=25$, even thinking in GIGAFLOPS, the task is not easy, since it will
be necessary at least $\Delta t\simeq $ $10^{12}$.$10^{-9}$ $=10^{3}\ $%
seconds , or $10^{3}/(60)$ $\simeq $ $17$ minutes which is only a rough
estimate. So the task gets more and more complicated even in faster
computers. And there is not point in arguing that in faster computers will
be possible to advance much more. For example, in order to enumerate the
configurations of a map with $6$ $\times $ $6=36$ countries in the same time
that we today perform the computation of the configurations for a map $5$ $%
\times $ $5$ it will be necessary to build a computer $170000$ faster than
these which we are considering. Well, but the things are no so bad!

%TCIMACRO{\TeXButton{B}{\begin{table}[tbp] \centering}}%
%BeginExpansion
\begin{table}[tbp] \centering%
%EndExpansion
\begin{tabular}{lllll}
\hline\hline
$N$ & $\Omega _{\text{colours}}^{(FBC)}$ & Time & $\Omega _{\text{colours}%
}^{(PBC)}$ & Time \\ \hline\hline
\multicolumn{1}{c}{4} & \multicolumn{1}{c}{18} & \multicolumn{1}{c}{$%
<O(10^{-2})$} & \multicolumn{1}{c}{18} & \multicolumn{1}{c}{$<O(10^{-2})$}
\\ 
\multicolumn{1}{c}{9} & \multicolumn{1}{c}{246} & \multicolumn{1}{c}{$%
<O(10^{-2})$} & \multicolumn{1}{c}{24} & \multicolumn{1}{c}{$<O(10^{-2})$}
\\ 
\multicolumn{1}{c}{16} & \multicolumn{1}{c}{7812} & \multicolumn{1}{c}{$%
\approx \ 0.17$ sec} & \multicolumn{1}{c}{4626} & \multicolumn{1}{c}{$%
\approx 0.18\ $sec} \\ 
\multicolumn{1}{c}{25} & \multicolumn{1}{c}{580986} & \multicolumn{1}{c}{$%
\approx 41$ min} & \multicolumn{1}{c}{38880} & \multicolumn{1}{c}{$\approx
39 $ min} \\ \hline\hline
\end{tabular}%
\caption{Time spent to enumerate the acceptable configurations to paint a map with 9 countries 
by the "brute force" method}\label{Table:Exact_BF}%
%TCIMACRO{\TeXButton{E}{\end{table}}}%
%BeginExpansion
\end{table}%
%EndExpansion

We already have in hands some exact results from small systems using the
\textquotedblleft brute force\textquotedblright\ method (see Table \ref%
{Table:Exact_BF}). In that table, we show the execution of the algorithm
using FBC and PBC. We also present the time required for the execution by
using a processor Intel i7-8565 U for both situations. Sure, the time
depends a lot on situations and we present the results of one execution only
for an idea of the order of magnitude. Such times are impracticable since
from $N=16$\ to $N=25$, which are very small systems, the time changes from
fraction of$\ $seconds approximately to spent times about hours. Thus, it is
mandatory to find a numerical alternative for working with larger systems
which permits to us to do an extrapolation to the thermodynamic limit: $%
N\rightarrow \infty $. This will be performed in the next section.

\section{Elegant numerical results: The transfer-matrix method}

\label{Section: Numerical_and_transfer}

For larger systems one has to resort a strategy very useful in Statistical
Physics, which reduces a two-dimensional problem to a succession of
one-dimensional problems. This approach is particularly useful for problems
with cylindrical or toroidal geometry \cite{Barber}, infinite in one of the
directions. The first attempts to work with finite systems in two directions
were made by Binder \cite{Binder}. However, it was Creswick \cite{Creswick}
who obtained in 1995 the most efficient way to apply the technique to this
geometry. Herein, we present this procedure in a system which is probably
the most simple case where the technique can be applied, without requiring
any prior knowledge of magnetic models and Statistical Mechanics. For that,
we used functions of the programming language Fortran (similar functions
exist in C language) and the task involves passing from one line (of
countries) to the next. In the context of computer science this type of
algorithm is known as greedy, since it discards the previously analysed
instances. To apply this technique we also use PBC in one of the directions
exactly as Creswick, which leads to a faster convergence at limit $%
N\rightarrow \infty $ when compared with FBC.

The calculation begins by choosing the possible colourings (configurations)
for any line of countries. A country can be coloured with one of the three
colours (from now on substituted by numbers $0,1$, and $2$) and two adjacent
countries cannot have the same colour (the last country cannot have the
colour of the first one due to PBC in the horizontal direction). Since $L$
is the width of the map there are 
\begin{equation*}
N_{\max }=(3-1)^{L}+(-1)^{L}(3-1)=2^{L}+2(-1)^{L}
\end{equation*}%
possible different colourings to colour a line of countries, as shown in Eq. %
\ref{Eq: Coloring_circuit} in the Appendix A.

%TCIMACRO{\TeXButton{B}{\begin{table}[tbp] \centering}}%
%BeginExpansion
\begin{table}[tbp] \centering%
%EndExpansion
\begin{tabular}{lll}
\hline\hline
$N$ & $\Omega _{\text{colours}}^{(PBC)}$ & Time\ (secs) \\ \hline\hline
4 & $18$ & $<O(10^{-2})$ \\ 
9 & $24$ & $<O(10^{-2})$ \\ 
16 & $4626$ & $<O(10^{-2})$ \\ 
25 & $38880$ & $<O(10^{-2})$ \\ 
36 & $37284186$ & $<O(10^{-2})$ \\ 
49 & $1886476032$ & $<O(10^{-2})$ \\ 
64 & $9527634436194$ & $\approx 0.016$ \\ 
81 & $2825260002442752$ & $\approx 0.047$ \\ 
100 & $77048019386428981200$ & $\approx 0.14$ \\ 
121 & $132046297983569476000000$ & $\approx 0.66$ \\ 
144 & $19698820973096973600000000000$ & $\approx 2.06$ \\ 
169 & $193554351965523488000000000000000$ & $\approx 12$ \\ 
196 & $159147870862109172000000000000000000000$ & $\approx 64$ \\ 
225 & $8920091695709351210000000000000000000000000$ & $\approx 333$ \\ 
\hline\hline
\end{tabular}%
\caption{Results obtained with the transfer matrix method. The
saving in processing times using this method when compared with  BF method
is notorious. One can run in a few minutes colourings with 225 countries.
From $N=100$ onwards, the zeros are placed only to complete the power obtained in the
numerical result since one has at most 18 significant digits in double
precision.} \label{Table:PBC}%
%TCIMACRO{\TeXButton{E}{\end{table}}}%
%BeginExpansion
\end{table}%
%EndExpansion

\begin{figure}[h]
\begin{center}
\includegraphics[width=1.0\columnwidth]{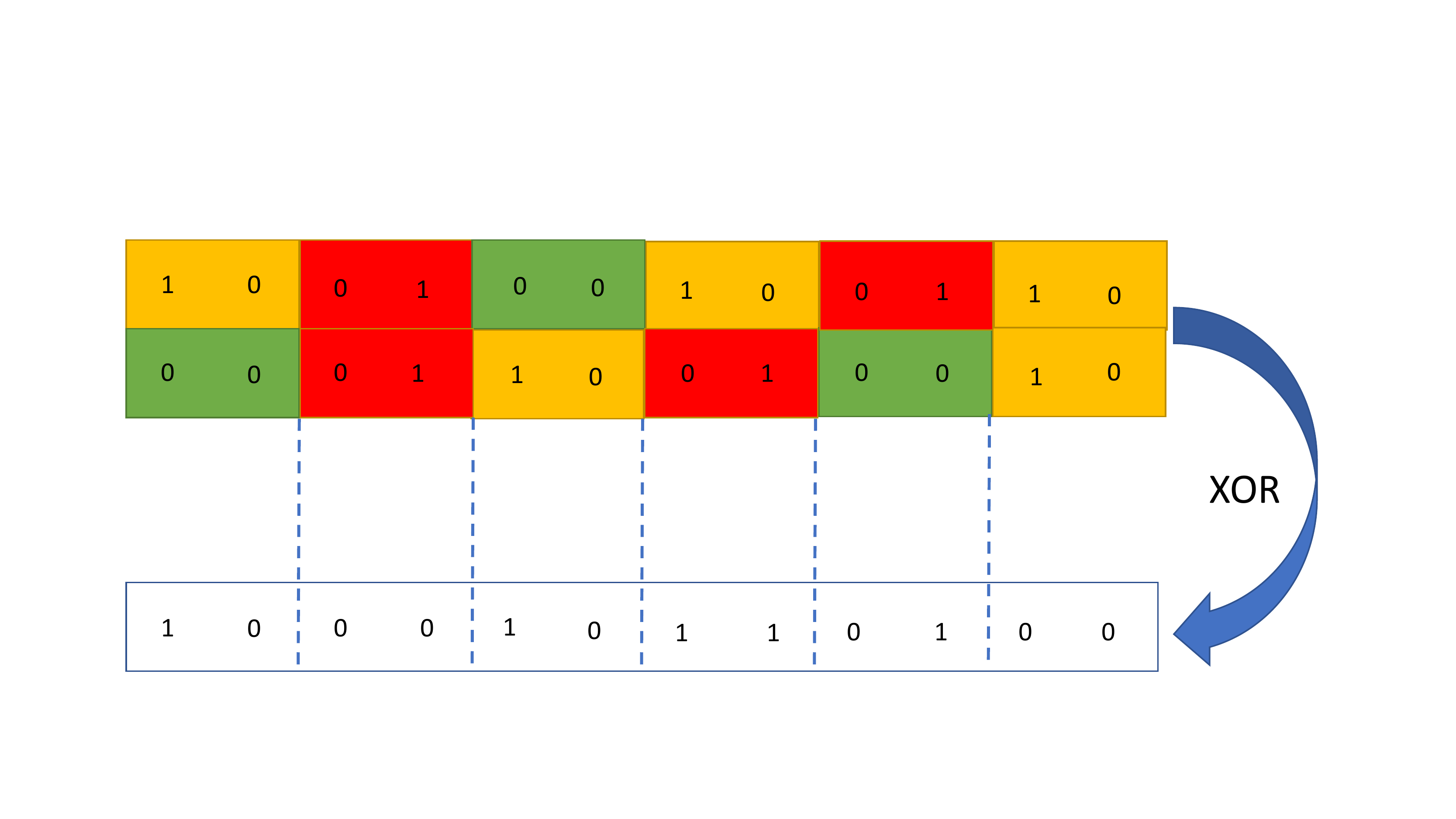}
\end{center}
\caption{Exclusive OR (XOR) operation between two configurations. A double
zero occurs only if one has the same bits occupying the same position in the
lines. }
\label{Fig:XOR}
\end{figure}

There are $L$ loops\textit{\ }and\textit{\ }($L/2+1$)\textit{\ }decision
commands (if's)\textit{\ }to discover the possible configurations during
evolution. Next, we move to the second line that can also present only one
of these acceptable configurations in the initial step. Among them, we need
to discover which are the compatible colourings with each configuration of
the first line, i.e., those with no two countries in the same position,
coloured with the same colour. This operation can be performed associating
each configuration of a line to an integer number, using a binary language.
This integer has $2L$ bits since each country needs two bits to store its
colour ($00$ corresponds to the colour $0$, $01$ corresponds to colour $1$,
and $10$ to colour $2$). We then apply the operation \textit{exclusive} $OR$
or simply $XOR$ ( $IEOR$ for Fortran compilers) to the two integers
representing the configurations of the first and second lines. Since XOR
(exclusive OR) works on all bits (see Fig. \ref{Fig:XOR}), then the result
is $0$ if the bits are equal in the same position, and 1 if they are
different. A double zero occurs only if one has the same bits occupying the
same position in the lines, corresponding to two countries with the same
colour.

\begin{figure}[h]
\begin{center}
\includegraphics[width=1.0\columnwidth]{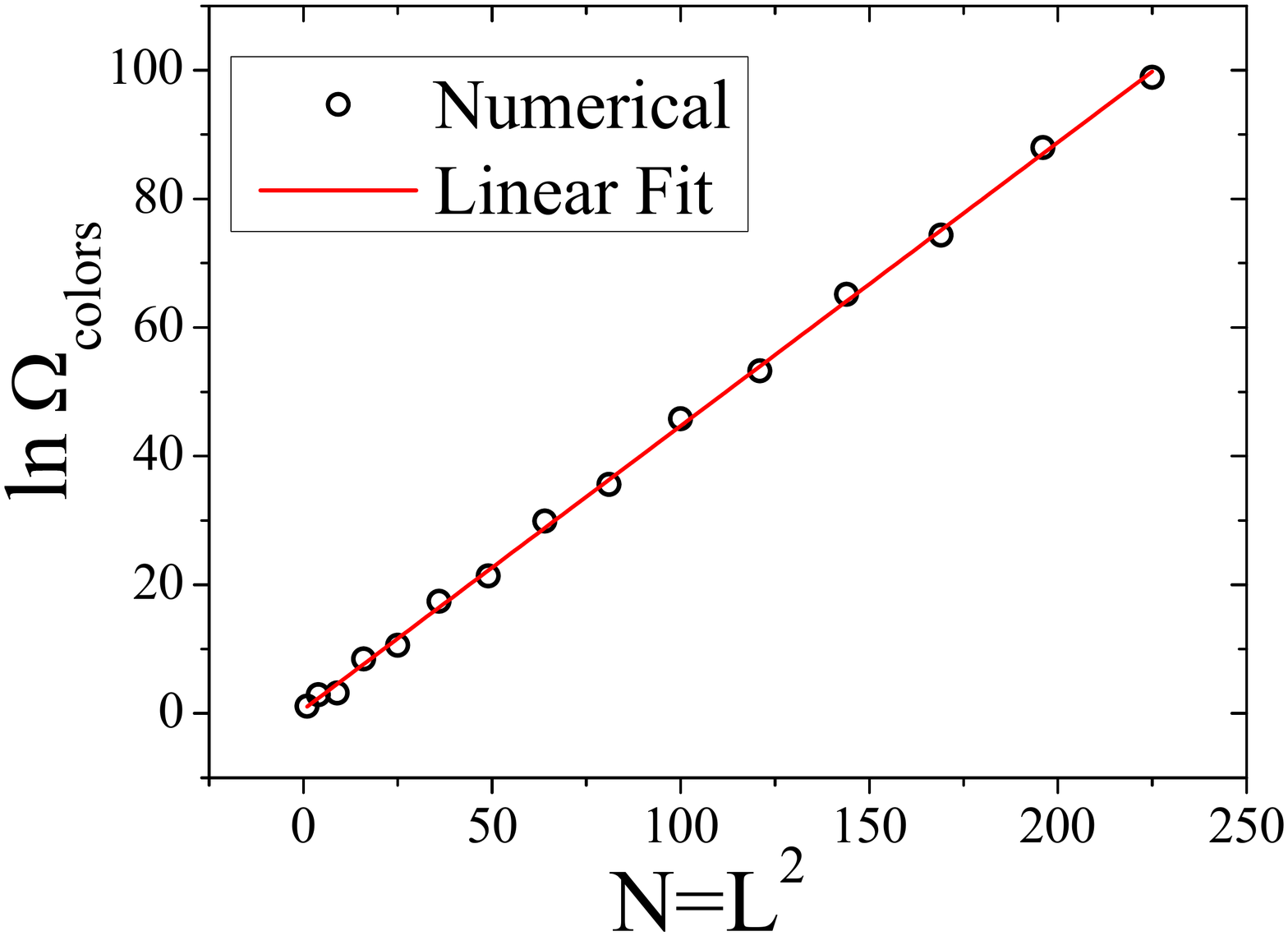}
\end{center}
\caption{Plot of $\ln \Omega $ obtained by the transfer matrix method as
function of $N$. It is expected that the slope of this curve gives an
estimate to $\ln W$. }
\label{Fig:Extrapolation}
\end{figure}

Thus, $L$ decision commands (if's) between bits of the integer resulting
from operation \textit{IEOR}(\textit{Line1,Line2}) are enough to detect the
existence of adjacent countries with the same colour. In this case, a
configuration of the second line will not be compatible with the first line.
After performing this selection, a comparison is made between the second and
third lines and so on until the last line is considered, since the PBC
condition does not apply in this direction. This is important to make the
algorithm faster; it is thus greedy, i.e., information of past lines is
discarded by using the subroutine \texttt{Transfere} accessed from the main
algorithm (shown in Appendix II of this paper for $N=225$). The results with
this algorithm in Table \ref{Table:PBC} are the same obtained by direct
counting (BF) in the simpler cases ($N=9,16$ and $25$) but now it can be
readily extended to $N=225$. It is then possible to perform an extrapolation
to $N\rightarrow \infty $, shown in Fig. \ref{Fig:Extrapolation}. According
to Eq. \ref{Eq:relation_omega_e_W} it is expected that $\ln \Omega _{\text{%
colours}}=\ln 3+N\ln W$, and a plot of $\ln \Omega _{\text{colours}}$ as
function of $N$ suggests a linear behaviour with a slope numerically equal
to $\ln W$, which is exactly observed in the Fig. \ref{Fig:Extrapolation}.

A linear fitting leads to $W_{num}=1.5421\pm 0.0054$ which must be compared
with the result obtained by Lieb in 1967 \cite{Lieb}:

\begin{equation*}
W=(4/3)^{(3/2)}\approx 1.5396007.
\end{equation*}%
Additional points about this result can be observed in Appendix I. After
this problem treated by Lieb, actually based on the work of Lee and Yang 
\cite{LeeYang}, other works involving six-vertex models were explored in the
literature. The novelty about these problems is that vertices are not
equally probable (they have no the same weight) since they represent
energetically distinct situations. These models present phase transition
even in one dimension when one changes the temperature \cite{Salinas} since
they do not obey the Mermin-Wagner theorem. But this is another history for
other opportunity!\ 

\section{Conclusions}

\label{Section: conclusions}

The problem of residual entropy of the ice was revisited by using a useful
procedure in Statistical Mechanics. To employ this method, it was necessary
to use a mapping of the ice-type model onto the problem of three colours.

The results with the transfer-matrix method indicate that even working with
small systems and considering periodic boundary conditions in only one
direction according to Creswick ideas, it is possible to obtain a good
estimate by performing an extrapolation to the thermodynamic limit of the
residual entropy. The use of the binary language in the representation of
the configurations and the possibility of comparing two configurations, by
using only simple operations with bits, is the other point that must be
highlighted.

We call the attention of the reader, mainly that one who is interested in
the Combinatorics and Number theory, for the exact result obtained by Lieb
for $W=\lim_{N\rightarrow \infty }\Omega ^{1/N}$ given by fraction of
integers (4/3) raised to (3/2) with $\Omega $ corresponding to the number of
the proper colourings of square lattices/maps with three colours. The
colouring with more colours of the same square lattices, on the other hand,
does not have an exact solution, only an upper and a lower bound are known
in the thermodynamic limit \footnote{%
For the interested readers, a brief review about graph colouring and other
important support results to this paper can be observed in Appendix I.} as
demonstrated by Biggs in 1977 \cite{Biggs77}.

Finally, the residual entropy of the ice is, no doubt, a interesting problem
of the Physics and our main pedagogical contribution is to show that the
transformation of the problem into the colouring map problem, makes possible
an elegant numerical solution, accessible by undergraduate students and with
only a basic knowledge in Mathematics and Computer Programming. In addition,
more interested readers can go beyond considering more technical aspects in
graph theory found in the first appendix, and that ones that desire to
explore the computer codes by using transfer matrix method.

\section*{Acknowledgments}

The authors would like to thank O. N. de Oliveira Jr. for his careful
reading of this manuscript and the priceless suggestions. R. da Silva thanks
CNPq for financial support under grant numbers 311236/2018-9, and
424052/2018-0.

\section*{Appendix I: Understanding a little bit about graph theory and the
problem of map colouring}

\label{Section:graph_theory}

\begin{figure}[h]
\begin{center}
\includegraphics[width=1.0\columnwidth]{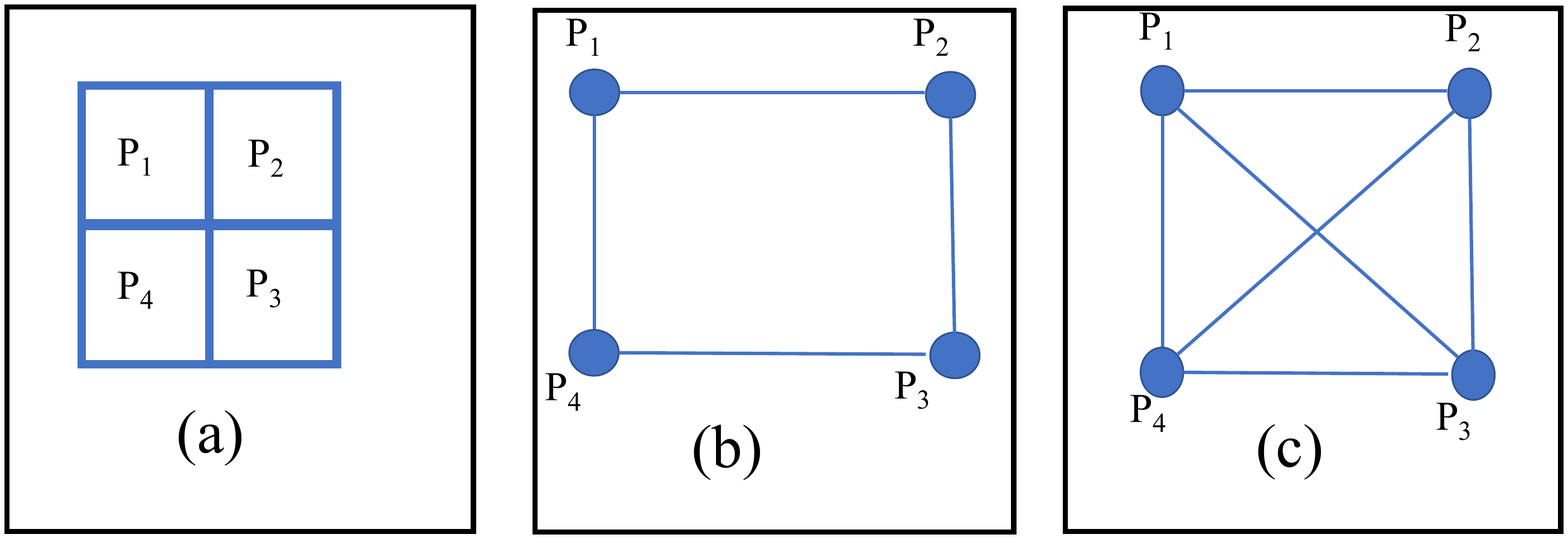}
\end{center}
\caption{Plot (a) Map with four countries. Plot (b) Corresponding graph of
the map represented in (a). Plot (c) This graph does not correspond to
situation (a) since the country $P_{1}$ is not a neighbor of $P_{3}$, and $%
P_{2}$ is not a neighbour of $P_{4}$. }
\label{Fig:maps_and_graphs}
\end{figure}

Looking from a graph theory point of view, we can observe maps as a set of
countries as vertices, while the borders between the countries representing
edges linking these vertices. We can observe a map with four countries and
its respective graph in the Figs. \ref{Fig:maps_and_graphs} (a) and \ref%
{Fig:maps_and_graphs} (b).

It is important to mention that the graph represented in Fig. \ref%
{Fig:maps_and_graphs} (c) does not correspond to the map observed in Fig. %
\ref{Fig:maps_and_graphs} (a) since the country $P_{1}$ is not a neighbour
of $P_{3}$, and $P_{2}$ is not a neighbour of $P_{3}$. Since we studied as
properly colouring with three colours a graph/map with four
vertices/countries, the Graph theory is much more general and we can extend
this for $x$ colours and even for more general graphs, which consists in a
interesting and important illustration for readers that want to understand a
little more about graph colouring.

In graph theory the number of ways to properly colour a graph with $x$
colours is so called the chromatic polynomial of this graph which is here
denoted by $\phi (x)$, and you will see that for the graph \ref%
{Fig:maps_and_graphs} (b) it is a easy step since you have understood the
case of 3 colours previously performed in this paper.

First there are only two possibilities: $P_{1}$ and $P_{3}$ have the same
colour, or they have different colours. In the first case, one has $x$ ways
to put the same colour simultaneously in $P_{1}$ and $P_{3}$ in this case
you can colour $P_{2}$ with $x-1$ colours while $P_{4}$ one also has $x-1$
ways since they are not neighbours, so one has in this first case

\begin{equation*}
\phi _{1}(x)=x(x-1)(x-1)=x(x-1)^{2}
\end{equation*}

On the other hand (second case) one has $P_{1}$ and $P_{3}$ painted with
different colours. There are $x(x-1)$\ ways to perform this task. For each
of these possibilities, one can paint $P_{2}$ with $x-2$ colours and $P_{4}$
also with $x-2$ colours, resulting in this second case a total number of
ways calculated by

\begin{equation*}
\phi _{2}(x)=x(x-1)(x-2)(x-2)=x(x-1)(x-2)^{2}
\end{equation*}

Thus the total number of colouring the graph \ref{Fig:maps_and_graphs} (b)
is 
\begin{equation}
\begin{array}{lll}
\phi (x) & = & \phi _{1}(x)+\phi _{2}(x) \\ 
&  &  \\ 
& = & x(x-1)^{2}+x(x-1)(x-2)^{2} \\ 
&  &  \\ 
& = & x(x-1)(x^{2}-3x+3)%
\end{array}
\label{Eq.four_sectors}
\end{equation}

A fast test of this formulae is to perform the particular case $x=3$, and
according to our previous calculations we must obtain $\phi (3)=18$, which
is exactly the result previously obtained. Please see the graph in Fig. \ref%
{Fig:maps_and_graphs} (c). In this case we have all vertices connected to
all vertices, and graphs that satisfy this condition are known as complete
graphs and are denoted by $K_{n}$. The particular case of Fig. \ref%
{Fig:maps_and_graphs} (c) corresponds to $K_{4}$ (complete graph with $n=4$
vertices) and it must be observed that only the number of vertices define
the graph since they have all possible edges (a total of $\binom{n}{2}$
edges for $K_{n}$). With some effort, one can observe that a proper
colouring of this graph demands $x\geq n$ colours. Thus, the chromatic
polynomial of the graph $K_{4}$ is easily calculated: 
\begin{equation*}
\phi _{K_{4}}(x)=x(x-1)(x-2)(x-3)
\end{equation*}%
and the result can be extended for $K_{n}$, using the fundamental counting
principle: 
\begin{equation*}
\phi _{K_{n}}(x)=x(x-1)(x-2)...(x-n+1)
\end{equation*}

\begin{figure}[h]
\begin{center}
\includegraphics[width=1.0\columnwidth]{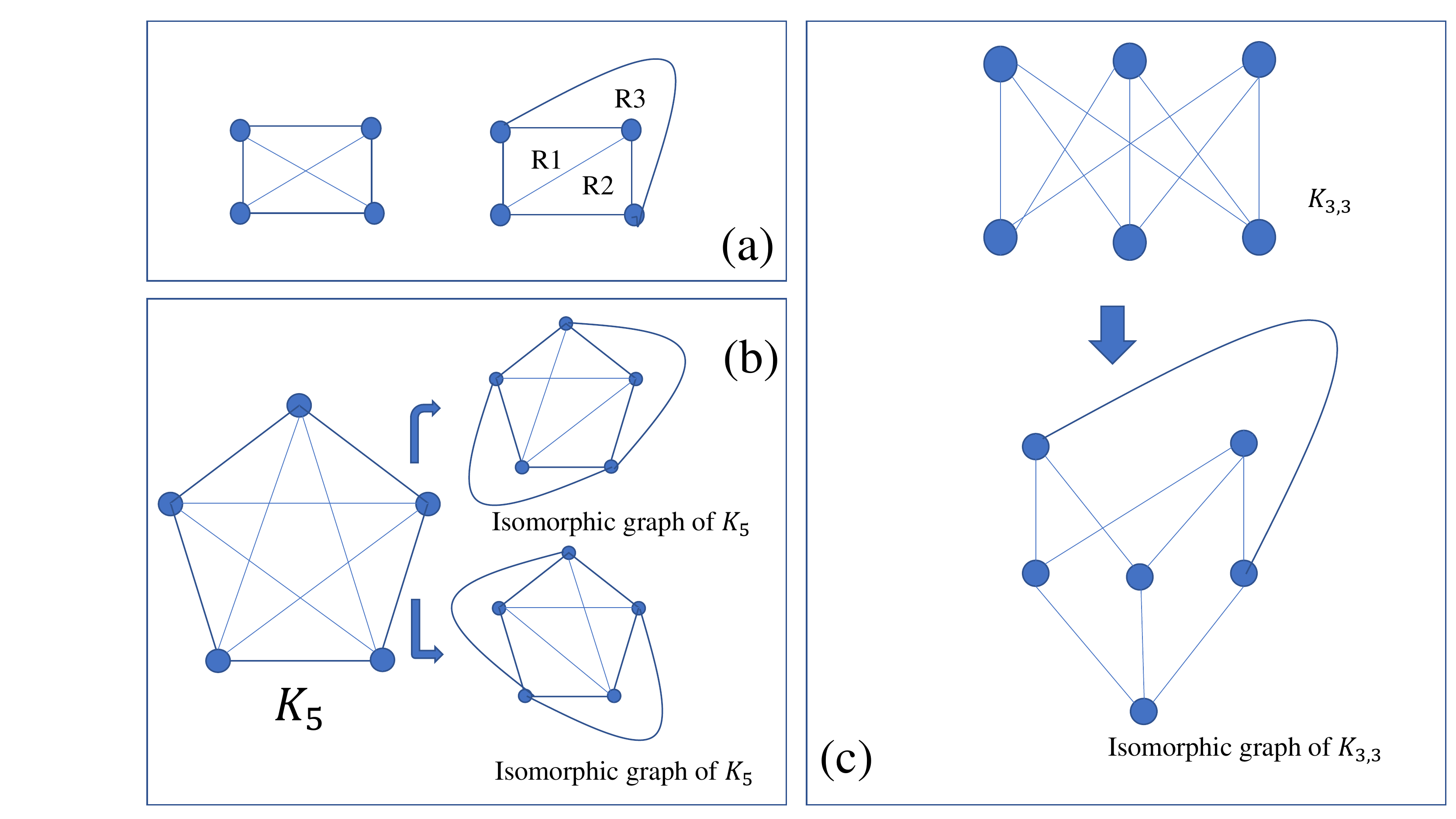}
\end{center}
\caption{(a) $K_{4}$ is a planar graph (b) $K_{5}$ is not a planar graph.
(c) $K_{3,3}$ is not also a planar graph. Both $K_{5}$ and $K_{3,3}$ are the
small graphs non planar and represent fundamental structures for the graph
theory. }
\label{Fig:planaridade}
\end{figure}

Actually, these graphs are much more \textquotedblleft sui
generis\textquotedblright\ than we can imagine, since for example for $n\geq
5$ they have not a planar representation (or in simple words, a map
representation), i.e., we cannot draw a planar representation of these
graphs without necessarily having two or more edges intersecting. Let us
better explain this point. For example, one observes that $K_{4}$ has a
notorious planar representation (see Fig. \ref{Fig:planaridade} (a)), we are
able to draw this graph as a map (in this case with 3 regions) or yet,
without the edges intersecting (of course, unless the vertices themselves).

On the other hand, we are not able to draw $K_{5}$ in a planar way. For
example in Fig. \ref{Fig:planaridade} (b) we observe two attempts (two
isomorphic graphs of $K_{5}$-- i.e., roughly speaking are the same graph $%
K_{5}$ drawn in a different way). No one of them leads to a planar
representation. The graph $K_{5}$ is a kind of \textquotedblleft minimal non
planar graph\textquotedblright . Other similar structure is the graph $%
K_{3,3}$ (complete bipartite graph on six vertices, three of which belonging
a set only connect to each of the three belonging to the other set) that can
be observed in Fig. \ref{Fig:planaridade} (c). Actually, any non-planar
graph cannot have a subgraph which is a subdivision of the $K_{5}$ or $%
K_{3,3}$. But why the planarity is a important concept if we are talking
about graph colouring? Because a fundamental theorem says that any planar
graph can be properly coloured with a maximum of 4 colours. The theorem was
demonstrated for the first time by 1976 by Kenneth Appel e Wolfgang Haken
(see for example \cite{Four-Color}), by using an IBM 360, the first accepted
proof by using a computer. For example $\phi
_{K_{5}}(x)=x(x-1)(x-2)(x-3)(x-4)$. If we make $x=4$, $\phi _{K_{5}}(4)=0$
which corroborates the theorem, since $K_{5}$ is a non-planar graph. On the
other hand $\phi _{K_{4}}(3)=0$, but $\phi _{K_{4}}(4)=24$ ways, which also
corroborates the theorem since $K_{4}$ is a planar graph.

\begin{figure}[h]
\begin{center}
\includegraphics[width=1.0\columnwidth]{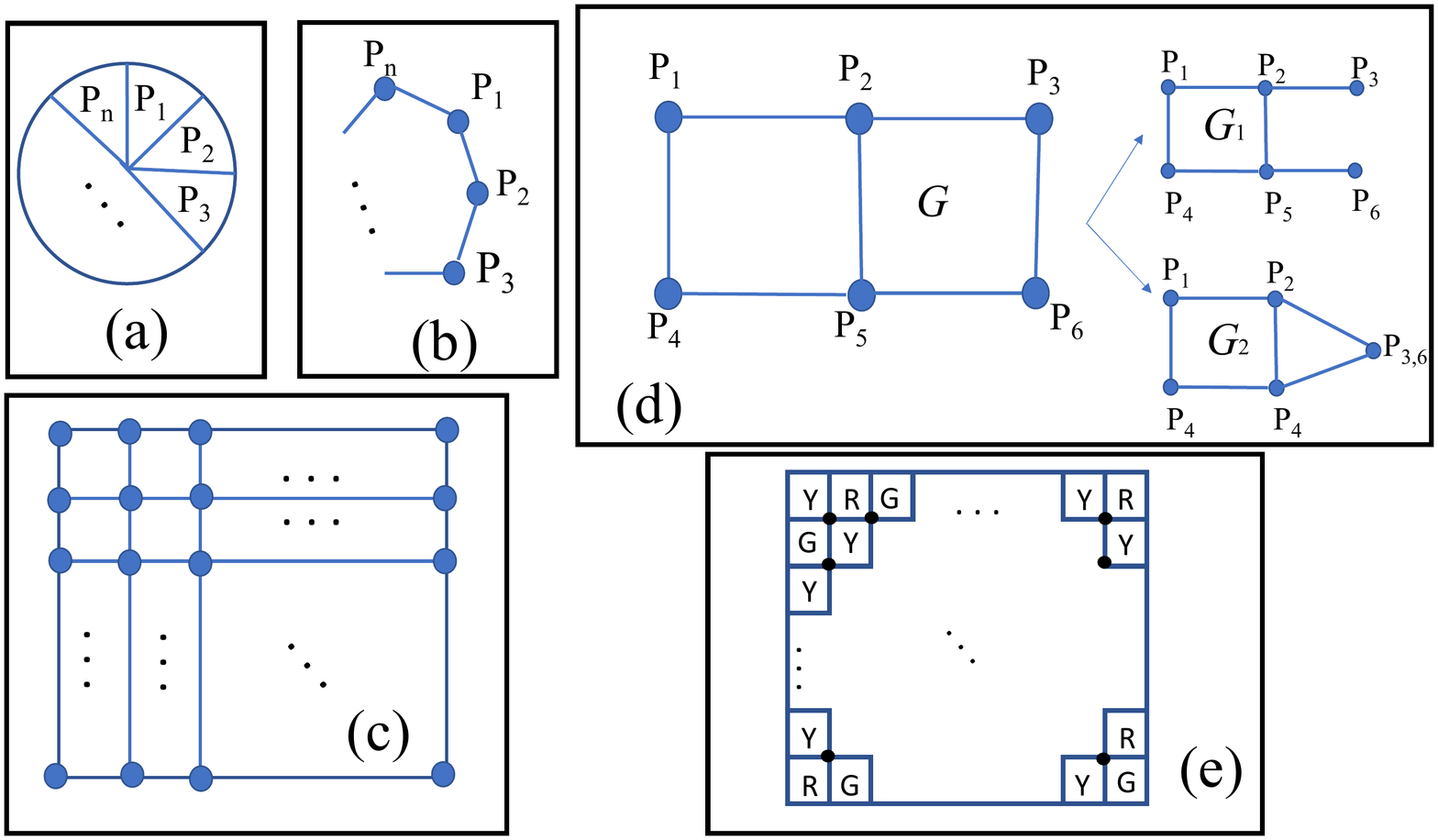}
\end{center}
\caption{(a) Circular sector, (b) graph representation of a circular sector
(c) A generalization of map with $N=L^{2}$ countries (a two-dimensional
lattice) (d) Graph corresponding a world with 6 countries (e) A particular
configuration of colours Y (yellow), G (green), and R (red) to the vertex of
the lattice represented in (c) where the blue balls are changed by square
cells (countries).}
\label{Fig:countries_and_graphs_2}
\end{figure}

Let us go back to our world with four countries. As we saw, it is more
complicated to colour this map than the graph $K_{4}$. We also can observe
that a world with four countries is a particular case of colouring a disk of 
$n$ sectors/countries (Fig. \ref{Fig:countries_and_graphs_2} (a) ), where
the Fig. \ref{Fig:countries_and_graphs_2} (b) is a graph representation of
the this world where the countries are placed as disk sectors. This graph is
known as a circular disk.

The idea is the same, for example, fixing two countries $P_{1}$ and $P_{3}$,
or any couple of countries (non-adjacent sectors) separated by only a sector
or neighbour a common sector (in this particular choice, $P_{2}$). In this
situation we have two options, that these countries ($P_{1}$ and $P_{3}$)
can be coloured with the same colour (situation I) or with two different
colours (situation II). Denoting $\phi _{n}(x)$ the number of ways to
properly colour this sector can be described by the recurrence relation 
\begin{equation}
\phi _{n}(x)=(x-2)\phi _{n-1}(x)+(x-1)\phi _{n-2}(x)
\label{Eq.recurrence_relation}
\end{equation}%
which demands some explanation. In situation I, i.e., $P_{1}$ and $P_{3}$
have the same colour works as if these two sectors were merged in a same
sector. Thus, for each colouring of the disk of $n-2$ sectors composed by
the sector originated from the fusion of the sector $P_{1}$ with sector $%
P_{3}$ and by other all sectors $n-3$ sectors (except by sector $P_{2}$),
one has $x-1$ ways to paint the sector $P_{2}$ which cannot have the same
colour of $P_{1}$ neither $P_{3}$. On the other hand (situation II), we have
that for each colouring of a disc with $n-1$ sectors composed by all sectors
except by the sector $P_{2}$, and for each painting of this disk we have $%
x-2 $ options to the sector $P_{2}$ which necessarily has a colour different
of the colours attributed to neighbouring sectors $P_{1}$ and $P_{3}$, which
justify the recurrence relation \ref{Eq.recurrence_relation}.

An interesting answer for this recurrence relation is $\varphi
_{n}(x)=\alpha ^{n}$, by direct substitution one has:%
\begin{equation*}
\alpha ^{2}-(x-2)\alpha -(x-1)=0
\end{equation*}%
that has two distinct roots: $\alpha _{1}=p-1$ and $\alpha _{2}=-1$, and a
general solution is given by the linear combination: $\phi
_{n}(x)=A(x-1)^{n}+B(-1)^{n}$. Such equation requires two initial conditions
which we know. First a disk with two sectors has $\phi _{2}(x)=x(x-1)$ ways
to be coloured, since the colour attributed to $P_{1}$ necessarily have a
colour different of $P_{2}$. In a disk with 3 sectors, all of them are
neighbours, thus similarly $\phi _{2}(x)=x(x-1)(x-2)$. So with these two
initial conditions we can conclude that $A=1$ and $B=x-1$, which results in 
\begin{equation}
\phi _{n}(x)=(x-1)^{n}+(-1)^{n}(x-1)  \label{Eq: Coloring_circuit}
\end{equation}

It is important to mention that such equation recovers the map with four
countries (disk with four sectors) since $\phi
_{4}(x)=(x-1)^{4}+(-1)^{4}(x-1)=x(x-1)(x^{2}-3x+3)$, exactly as we obtained
in Eq. \ref{Eq.four_sectors}. After this tour across the graph theory and
its connection with the colouring of the graphs, let us come back to the
colouring of worlds with many countries. We already study the simple case of
world with four countries described by Fig. \ref{Fig:maps_and_graphs} (a)
and represented by Fig. \ref{Fig:maps_and_graphs} (b). In the case of many
countries which is represented by the two-dimensional lattice (Fig. \ref%
{Fig:countries_and_graphs_2} (c) ), the colouring is not easy. We can start
extending the world of four countries to six countries (Fig. \ref%
{Fig:countries_and_graphs_2} (d) ). \ An important theorem in graph theory
is the deletion-contraction theorem. This theorem says that for example
choosing the edge between the countries $P_{1}$ and $P_{6}$ in the original
graph ($G$), the chromatic polynomial of $G$ is the polynomial of the graph
obtained by deletion of this edge ($G_{1}$) minus the polynomial of the
graph obtained by contraction of this edge ($G_{2}$). To calculate $\phi
_{G_{1}}(x)$, we can observe that it is obtained multiplying $\phi _{4}(x)$
times the ways of properly colouring the vertex $P_{3}$, which occurs in $%
x-1 $ possible ways, times the ways of properly colouring the vertex $P_{6}$
which also occurs in $x-1$ possible ways, so:%
\begin{equation*}
\begin{array}{lll}
\phi _{G_{1}}(x) & = & \phi _{4}(x).(x-1)(x-1) \\ 
&  &  \\ 
& = & x(x-1)^{3}(x^{2}-3x+3)\text{.}%
\end{array}%
\end{equation*}%
On the other hand, we must observe that the vertex $P_{3,6}$ has a stronger
restriction, it can be coloured with colours different of $P_{2}$ and $P_{4}$
that always have different colours, thus 
\begin{equation*}
\begin{array}{lll}
\phi _{G_{2}}(x) & = & \phi _{4}(x)(x-2) \\ 
&  &  \\ 
& = & x(x-1)(x-2)(x^{2}-3x+3)%
\end{array}%
\end{equation*}

So, one has 
\begin{equation*}
\begin{array}{lll}
\phi _{G}(x) & = & \phi _{G_{1}}(x)-\phi _{G_{2}}(x) \\ 
&  &  \\ 
& = & x(x-1)^{3}(x^{2}-3x+3)-x(x-1)(x-2)(x^{2}-3x+3) \\ 
&  &  \\ 
& = & x(x-1)(x^{2}-3x+3)^{2}%
\end{array}%
\end{equation*}

With $x=3$ colours, we obtain $\phi _{G}(3)=54$ ways. In a more general
case, one can attribute three colours (the same previous colours: Yellow,
Green, and Red) to the vertices in the lattice (Fig. \ref%
{Fig:countries_and_graphs_2} (c) ) such that neighbouring vertices cannot
have the same colour, exactly as countries in a map, by following our
convention where vertices (little balls) correspond to cells (countries), as
for example we can observe in Fig. \ref{Fig:countries_and_graphs_2} (e).We
should naively imagine that recursively an expression for the lattice with $%
N=L^{2}$ countries should be obtained. But is is not true! Actually we have
no an analytical expression for an arbitrary $N$.

In \cite{Biggs77} for example it is shown that for $N\rightarrow \infty $
upper and lower bounds are obtained:%
\begin{equation*}
\frac{1}{2}(x-2+\sqrt{x^{2}-4x+8})\geq \phi (x)\geq \frac{x^{2}-3x+3}{x-1}
\end{equation*}

However for $x=3$, both bounds are the golden ratio $\frac{1}{2}(1+\sqrt{5})$
and $3/2$. However Lieb in a brilliant work has obtained a exact result for $%
x=3$ at limit $N\rightarrow \infty $: $\phi _{\infty }(3)=(\frac{4}{3}%
)^{3/2} $. But is $x=3$ an important case to us? Absolutely, since we can
show that three-colouring problem is exactly the six-vertex problem
(ice-type model) except by a multiplicative factor, which is exactly our
original problem as we show in the the next subsection.

\newpage

\section*{Appendix II: Transfer matrix algorithm for L=15 using PBC in one
of the directions}

\label{Section:Algorithm_transfer_matrix}

1. \textbf{Main algorithm}

\begin{center}
\includegraphics[width=14.7cm]{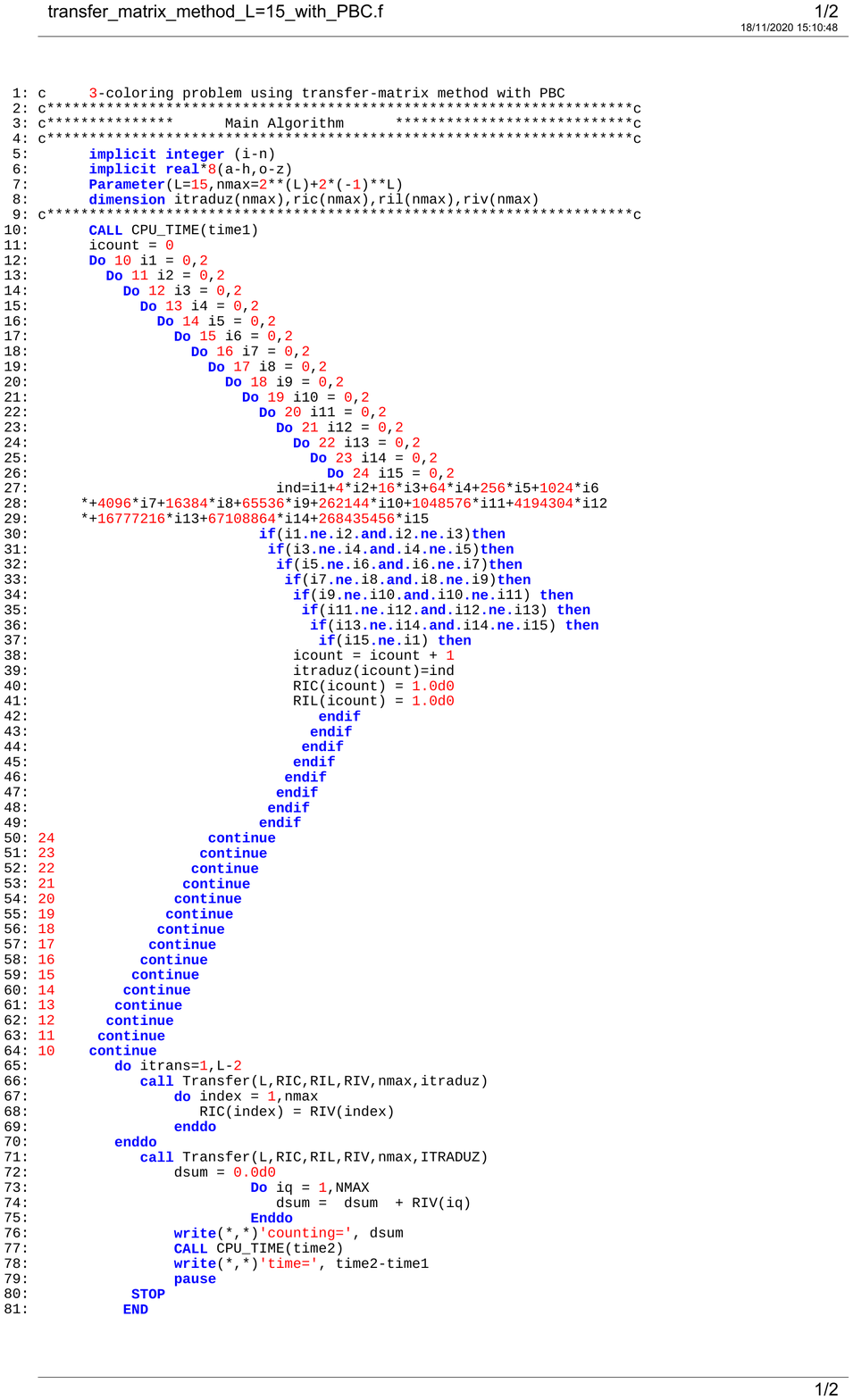}
\end{center}

\newpage

2. \textbf{Subroutine}

\begin{center}
\includegraphics[width=16cm]{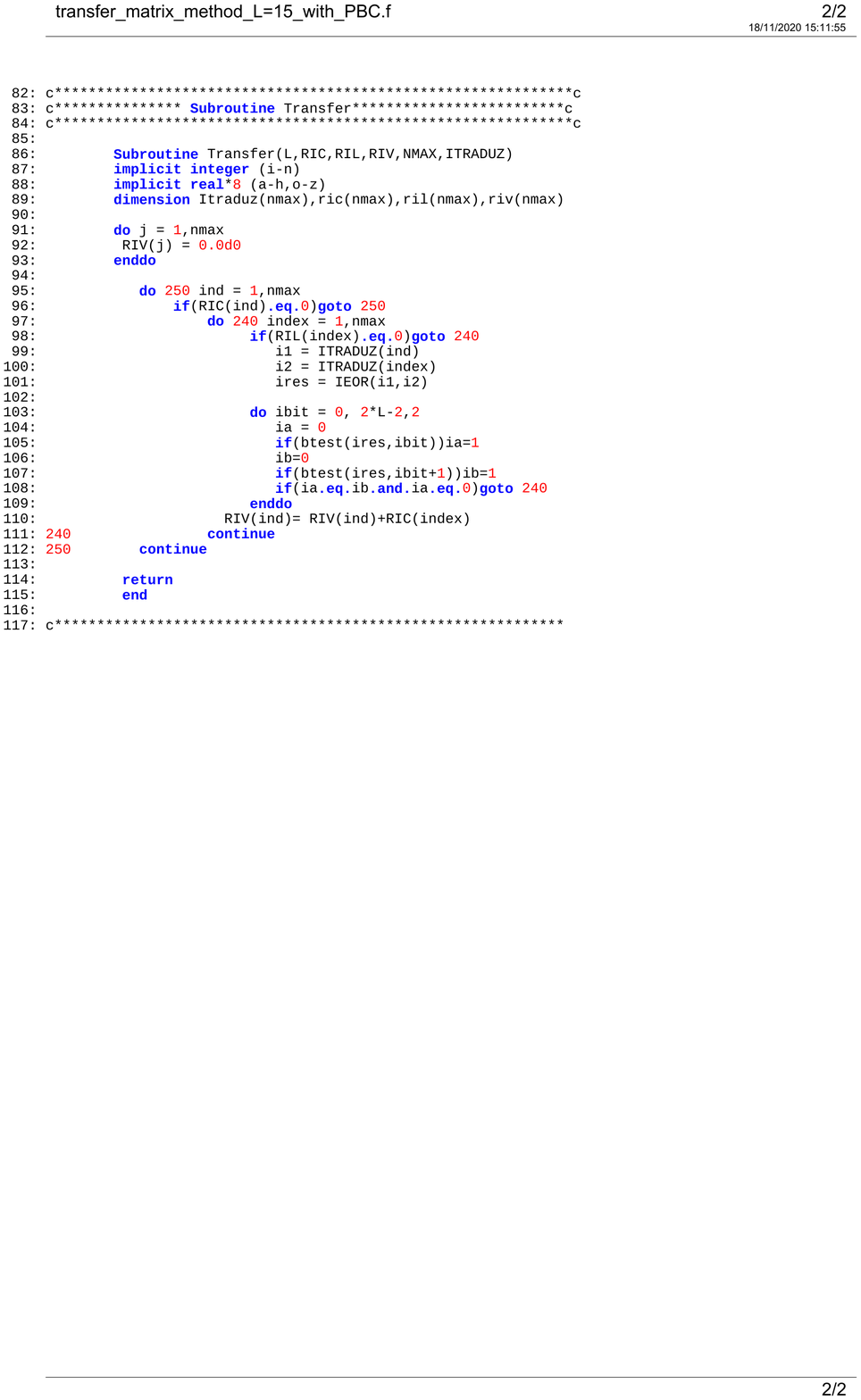}
\end{center}


\begin{thebibliography}{99}
\bibitem{Nagle2} It is important to mention that Nagle presented the
estimate $W=1.540\pm 0.001$ in his original work and not exactly an estimate
for the entropy of the two-dimensional version of the ice-type model model.
We take the liberty to present the estimate for $s$ based on this value
considering that the uncertainty in $s$ was performed via error propagation: 
$\sigma _{s}=\sqrt{\ln ^{2}W\cdot \sigma _{k_{B}}^{2}+\frac{k_{B}^{2}}{W^{2}}%
\sigma _{W}^{2}}$ $\approx \frac{k_{B}}{W}\sigma _{W}$, once $\sigma
_{W}>>\sigma _{k_{B}}$, where $\sigma _{k_{B}}$ and $\sigma _{W}$ are
respectively the uncertainties in $k_{B}$ and $W$, since $%
k_{B}=(1.9872041\pm 0.0000018)$ cal/mol K.
\end{thebibliography}
\end{document}